\documentclass[12pt,english,draftcls, one column]{IEEEtran}
\usepackage[T1]{fontenc}
\usepackage[latin9]{inputenc}
\usepackage{float}
\usepackage{amsmath}
\usepackage{amsthm}
\usepackage{amssymb}
\usepackage{graphicx}
\usepackage{multirow}
\usepackage{xcolor}
\usepackage{subcaption}
\usepackage{multicol}
\usepackage{diagbox}
\usepackage{multirow}
\usepackage{cite}
\usepackage{algorithm, algorithmic}

\makeatletter

\floatstyle{ruled}
\newfloat{algorithm}{tbp}{loa}

\floatname{algorithm}{\protect Algorithm}

\theoremstyle{plain}

\theoremstyle{plain}

\ifCLASSINFOpdf
\else
\fi
\usepackage{babel}

\makeatother

\usepackage{babel}
\providecommand{\propositionname}{Proposition}
\providecommand{\theoremname}{Theorem}

\begin{document}


\title{Learning Decentralized Power Control in Cell-Free Massive MIMO Networks}

\author{Daesung Yu, \textit{Student Member}, \textit{IEEE}, Hoon Lee, \textit{Member}, \textit{IEEE}, \\Seung-Eun Hong, and Seok-Hwan Park, \textit{Member}, \textit{IEEE}\vspace{-10mm}\thanks{


This work was supported in part by National Research Foundation (NRF) of Korea Grants funded by the Ministry of Education under Grant 2019R1A6A1A09031717 and Grant 2021R1A6A3A13046157; in part by the Korea government Ministry of Science and ICT (MSIT) under Grant 2021R1C1C1006557 and Grant 2021R1I1A3054575; in part by Institute of Information \& communications Technology Planning \& Evaluation (IITP) grant funded by the MSIT (5G Open Intelligence-Defined RAN (ID-RAN) Technique Based on 5G New Radio) under Grant 2018-0-01659 and (Intelligent 6G Wireless Access System) under Grant 2021-0-00467.

D. Yu and S.-H. Park are with the Division of Electronic Engineering and the Future Semiconductor Convergence Technology Research Center, Jeonbuk
National University, Jeonju, Korea (email: \{imcreative93, seokhwan\}@jbnu.ac.kr).

H. Lee is with the Department of Information and Communications Engineering, Pukyong National University, Busan, Korea (email: hlee@pknu.ac.kr).

S.-E. Hong is with the Future Mobile Communication Research Division, Electronics and Telecommunications Research Institute, Daejeon 34129, South Korea (email: iptvguru@etri.re.kr).

Copyright (c) 2015 IEEE. Personal use of this material is permitted. However, permission to use this material for any other purposes must be obtained from the IEEE by sending a request to pubs-permissions@ieee.org.}}
\maketitle
\begin{abstract}
This paper studies learning-based decentralized power control methods for 
cell-free massive multiple-input multiple-output (MIMO) systems where a central processor (CP) controls access points (APs) through fronthaul coordination. To determine the transmission policy of distributed APs, it is essential to develop a network-wide collaborative optimization mechanism. 
To address this challenge, we design a cooperative learning (CL) framework which manages computation and coordination strategies of the CP and APs using dedicated deep neural network (DNN) modules. To build a versatile learning structure, the proposed CL is carefully designed such that its forward pass calculations are independent of the number of APs. To this end, we adopt a parameter reuse concept which installs an identical DNN module at all APs. Consequently, the proposed CL trained at a particular configuration can be readily applied to arbitrary AP populations. Numerical results validate the advantages of the proposed CL over conventional non-cooperative approaches.
\end{abstract}
\theoremstyle{theorem}
\newtheorem{theorem}{Theorem} 
\theoremstyle{proposition}
\newtheorem{proposition}{Proposition} 
\theoremstyle{lemma}
\newtheorem{lemma}{Lemma} 
\theoremstyle{corollary}
\newtheorem{corollary}{Corollary} 
\theoremstyle{definition}
\newtheorem{definition}{Definition}
\theoremstyle{remark}
\newtheorem{remark}{Remark}

\section{Introduction}
Cell-free massive MIMO has been regarded as a promising solution for the next-generation wireless networks owing to the enhanced coverage. 
In the cell-free massive MIMO setup, a central processor (CP) manages the transmission strategies of distributed access points (APs) through fronthaul links. To enhance the system performance, classical optimization approaches rely on centralized signal processing at the CP by collecting channel state information (CSI) measured at the APs \cite{Choi-et-al:TWC20}. However, this poses prohibitive fronthaul signaling overheads to frequently update short-term CSI. To address this issue, long-term CSI-based power control schemes were presented in \cite{Nayebi-et-al:TWC17, Ngo-et-al:TWC17} where the transmit power levels of all APs are centrally optimized at the CP by using large-scale fading state only. Due to the absence of the short-term CSI, a heuristic approximation of the performance measure, e.g., the achievable data rate, is adopted as the objective function. This brings a model mismatch between the actual system performance and its approximation, posing an optimality loss.

This issue can be handled by the data-driven optimization capability of the deep learning (DL) techniques.
There have been recent works on DL-based solutions to cell-free massive MIMO systems \cite{Yu-et-al:WCL21, Zaher-et-al:arXiv21, Lee-et-al:TWC21}. A joint optimization task of fronthaul quantization and multi-antenna beamforming was addressed in \cite{Yu-et-al:WCL21}. An ideal case is assumed where the CP perfectly knows channel vectors acquired at the APs. The CP employed a centralized deep neural network (DNN) to produce optimized solutions using the global network CSI. Such a centralized DL structure is not practical due to the high signaling overhead in the fronthaul coordination. This issue can be resolved via a decentralized DNN architecture \cite{Yu-et-al:WCL21, Zaher-et-al:arXiv21, Lee-et-al:TWC21}. The work in \cite{Zaher-et-al:arXiv21} realized decentralized power control schemes with the aid of multiple DNN modules at individual APs. To enhance the decentralized decision-making process, a cooperative learning (CL) strategy is an essential requirement to allow the APs to share their locally available statistics, e.g., local CSI. A heuristic coordination policy was developed in \cite{Zaher-et-al:arXiv21}, where the CP broadcasts man-made control messages to the APs. Such a synthetic CL strategy is, however, not optimal for DNN modules. To address this issue, \cite{Lee-et-al:TWC21} employed an additional DNN at the CP to generate self-organizing coordination messages autonomously. The CP DNN is trained along with the AP DNNs to maximize the desired objective function. As a result, both the decentralized solution calculation as well as fronthaul coordination rules are optimized jointly, thereby leading to improved system capacity. Existing DL approaches have fixed computation structures dedicated to a particular cell-free massive MIMO network. This loses the scalability of the network size. For instance, the DNNs presented in \cite{Yu-et-al:WCL21, Zaher-et-al:arXiv21, Lee-et-al:TWC21} cannot be directly applied to other setups with different AP populations. For this reason, we need to prepare a number of DNNs trained for all possible network configurations.

This paper proposes a versatile CL framework that is adaptive to arbitrary given cell-free massive MIMO configurations, especially, randomly varying AP population. We maximize the ergodic sum-rate performance by optimizing transmit power allocation variables at the APs in a decentralized manner. We first propose an analytical optimization process termed by the cooperative stochastic gradient descent (CSGD) algorithm. Based on the sample-average-approximation (SAA) approach \cite{Shamir-Srebro:Allerton14}, this method allows each AP to learn an effective solution individually using the SGD method, but with iterative fronthaul cooperation. Therefore, albeit its effectiveness, the CSGD is not practical with limited fronthaul resources.

To this end, we present a novel CL framework which identifies an efficient power control solution via one-shot fronthaul coordination. Decentralized calculations of the CP and the APs are handled by individual DNN units. These component DNNs are responsible to generate communication messages shared among the CP and the APs as well as to decide transmit power levels at individual APs. To establish a universal CL structure that scales up with the AP populations, we employ the parameter reuse technique which leverages the identical DNN module across all APs. In addition, the architecture of the component DNNs is carefully constructed such that they can work with arbitrary input/output dimensions independent of the number of the APs. Consequently, the proposed CL can be universally applied to any given cell-free massive MIMO networks. Numerical results confirm the effectiveness of the proposed CL over existing DL methods.

\section{System model and Scenario\label{sec:System-Model}}

A cell-free massive MIMO system is considered, where a CP
manages $M$ single-antenna APs\footnote{Although we assume the single-antenna AP scenario for the compaction, extension to the multi-antenna AP scenario is straightforward. The impact of the multi-antenna AP is also addressed in Sec. \ref{sec:Numerical-Results}.} to communicate with
$K$ single-antenna user equipments (UEs). We define the index sets $\mathcal{K}\triangleq\{1,2,\ldots,K\}$ and $\mathcal{M}\triangleq\{1,2,\ldots,M\}$ of UEs and APs, respectively.
Let $h_{k,i}\sim\mathcal{CN}(0,\rho_{k,i})$ be the channel coefficient between AP $i$ and UE $k$ where $\rho_{k,i} \triangleq \mathbb{E}[|h_{k,i}|^2]$ is the long-term path-loss of the corresponding link.
By using the standard channel acquisition process, each AP $i$ obtains local CSI 
estimates $\hat{\mathbf{h}}_i \triangleq \{\hat{h}_{k,i}\}_{k\in\mathcal{K}}$ of the actual channel coefficients $\mathbf{h}_{i}\triangleq\{h_{k,i}\}_{k\in\mathcal{K}}$. Here, $\hat{h}_{k,i}$ is modeled as \cite{Choi-et-al:TWC20, Yin-et-al:JSAC13}
\begin{align}
    h_{k,i} = \hat{h}_{k,i} + e_{k,i}, \label{eq:additive-CSI-error-model}
\end{align}
where $e_{k,i}$ accounts for the estimation error. With the linear minimum mean squared error (LMMSE) estimator, $e_{k,i}$ is uncorrelated to the nominal channel $\hat{h}_{k,i}$. From \cite{Nosrat-Makouei-at-al:TSP11}, it has been known that $\hat{h}_{k,i}$ and $e_{k,i}$ follow the complex Gaussian distribution as
\begin{align}\label{eq:sample}
    \hat{h}_{k,i} \sim \mathcal{CN}(0, (1-\phi) \rho_{k,i}) \text{ and } e_{k,i} \sim \mathcal{CN}(0, \phi \rho_{k,i})
\end{align}
where $\phi\in[0,1]$ stands for the error ratio. The error ratio depends on the signal-to-noise ratio of pilot symbols \cite{Nosrat-Makouei-at-al:TSP11}. Thus, it can be regarded as a random number that dynamically varies according to the propagation environment.


There are a number of powerful transmission strategies developed for the cell-free massive MIMO systems, e.g., the zero-forcing (ZF) and the regularized ZF (RZF) beamforming methods \cite{Riera-Palou-et-al:Allerton19}. To realize these centralized interference management schemes, the APs should share their local CSI estimates $\{\hat{\mathbf{h}}_{i}\}_{i\in\mathcal{M}}$ with the CP via the fronthaul coordination. However, frequent updates of this short-term CSI incur prohibitive fronthauling overheads. One practical solution is to let each AP $i$ forward its local long-term CSI $\boldsymbol{\rho}_{i} = \{\rho_{k,i}\}_{k\in\mathcal{K}}$ to the CP \cite{Nayebi-et-al:TWC17,Ngo-et-al:TWC17}. By doing so, the CP can still mitigate the multi-user interference with large-scale fading while reducing the signaling overhead in the fronthaul coordination \cite{Dartmann-et-al:VTC09}.

For this reason, the decentralized beamforming policy is considered at individual APs to determine their beam weights by using the local short-term CSI only. In particular, the conjugate beamforming (CB) scheme is adopted which can be computed in a decentralized manner, thereby leading to the cost-effective fronthaul coordination \cite{Ngo-et-al:TWC17,Ngo-et-al:SPAWC15}.
The transmit signal $x_i$ of AP $i$ is given~as
\begin{align}
    x_i = \sum\nolimits_{k\in\mathcal{K}}{\sqrt{p_{k,i}} \, \frac{\hat{h}^*_{k,i}}{|\hat{h}^*_{k,i}|} \, s_k }, \,\,\, i\in\mathcal{M}, \label{eq:linear-conjugate-beamforming}
\end{align}
where $s_k$ denotes the data symbol intended for UE $k$ and $p_{k,i}$ represents the transmit power at AP $i$ allocated to UE $k$. 
The transmit power of AP $i$, defined as $\mathbf{p}_{i}\triangleq \{p_{k,i}\}_{k\in\mathcal{K}}$, is subject to the per-AP power budget $P$ as $\sum\nolimits_{k\in\mathcal{K}}p_{k,i}\leq P, \, i\in\mathcal{M}$.
The ergodic rate $R_k$ of UE $k$, which is averaged over a joint distribution of $(\hat{\mathbf{h}}, \mathbf{e}, \boldsymbol{\rho})$, is written as
\vspace{-1mm}
\begin{align}
    R_k = \mathbb{E}_{ \hat{\mathbf{h}}, \mathbf{e}, \boldsymbol{\rho} } \left[ \log_2\left( 1 + \gamma_k(\hat{\mathbf{h}}, \mathbf{e}, \mathbf{p})\right) \right], \label{eq:ergodic-rate-UE-k}
\end{align}
where $\mathbf{p}\triangleq\{\mathbf{p}_{i}\}_{i\in\mathcal{M}}$, $\mathbf{e}\triangleq \{e_{k,i}\}_{k\in\mathcal{K},i\in\mathcal{M}}$, and $\gamma_k(\hat{\mathbf{h}}, \mathbf{e}, \mathbf{p})$ stands for the signal-to-interference-plus-noise ratio (SINR) defined as
\begin{align}
    \gamma_k(\mathbf{\hat{h}},\mathbf{e},\mathbf{p}) \!=\! {\frac{\Big|\!\sum_{i\in\mathcal{M}}h_{k,i} \hat{h}_{k,i}^* \sqrt{p_{k,i}} / |\hat{h}_{k,i}^*| \Big|^2}{1 \!\!+\!\! \sum_{l\in\mathcal{K}\setminus{\{k\}}}\!\Big|\!\sum_{i\in\mathcal{M}}h_{k,i} \hat{h}_{l,i}^* \sqrt{p_{l,i}} / |\hat{h}_{l,i}^*| \Big|^2}}. \label{eq:SINR}
\end{align}

It is desired to maximize the average sum-rate performance by optimizing the transmit power $\mathbf{p}$ for each given realization $(\hat{\mathbf{h}},\mathbf{e},\boldsymbol{\rho})$. The corresponding problem is expressed as
\begin{subequations} \label{eq:problem-ergodic}
\begin{align}
&\underset{\mathbf{p}}{\mathrm{max.}}\,\, \mathbb{E}_{ \hat{\mathbf{h}}, \mathbf{e}, \boldsymbol{\rho} } \left[ \sum\nolimits_{k\in\mathcal{K}} \log_2\left( 1 + \gamma_k(\hat{\mathbf{h}}, \mathbf{e}, \mathbf{p})\right) \right] \label{eq:problem-ergodic-object}\\
&\,\,\mathrm{s.t.}\,\,\,\, \sum\nolimits_{k\in\mathcal{K}}p_{k,i}\leq P, \, \forall i\in\mathcal{M}.
\label{eq:power}
\end{align}
\end{subequations}
Problem \eqref{eq:problem-ergodic} is, in general, nonconvex, and thus it is not trivial to obtain the globally optimal solution. The expectation over arbitrary distributed CSI $(\hat{\mathbf{h}},\mathbf{e},\boldsymbol{\rho})$ has no analytical formula, which makes it difficult to apply traditional nonconvex optimization techniques. Conventional methods \cite{Nayebi-et-al:TWC17,Ngo-et-al:TWC17} have proposed tractable closed-form approximations for the average rate objective. To this end, all the small-scale fading coefficients are simply removed by using the Jensen's inequality, which brings the model mismatch between the ergodic rate and its approximated value. In addition, since the approximated rate expression depends only on the long-term channel statistics, there is no room for exploiting the short-term CSI in optimizing the power control variables.
Moreover, the separately deployed APs request a novel decentralized computation structure. Each AP $i$ should infer its local power allocation solution $\mathbf{p}_{i}$ based only on the partial network knowledge, i.e., the local CSI vectors $\hat{\mathbf{h}}_{i}$ and $\boldsymbol{\rho}_{i}$. Such a partial observation is insufficient to recover the optimal solution of \eqref{eq:problem-ergodic} individually. Thus, interaction among APs is essential to build efficient power control schemes.

\section{Cooperative Stochastic Gradient Descent}

This section presents a CSGD algorithm which handles the expectation term in \eqref{eq:problem-ergodic-object} using the SAA approach \cite{Shamir-Srebro:Allerton14}. The CSGD does not require the closed-form approximation for the sum-rate objective function. As a result, the CSGD algorithm provides an upper bound performance for existing power control schemes \cite{Nayebi-et-al:TWC17,Ngo-et-al:TWC17} that invokes model-mismatch errors. A key idea is to let each AP update its local solution $\mathbf{p}_{i}$ individually based on the standard mini-batch SGD algorithm. The APs first share their local CSI, in particular, the long-term CSI $\boldsymbol{\rho}_{i}$, via fronthaul links. From the known distribution in \eqref{eq:sample}, each AP $i$ can generate a mini-batch set $\mathcal{B}_{i}$ containing its local CSI error $\mathbf{e}_{i}\triangleq\{e_{k,i}\}_{k\in\mathcal{K}}$ as well as other-AP statistics $\hat{\mathbf{h}}_{-i}\triangleq\{\hat{h}_{k,j}\}_{j\in\mathcal{M}\backslash \{i\},k\in\mathcal{K}}$ and $\mathbf{e}_{-i}\triangleq\{e_{k,j}\}_{j\in\mathcal{M}\backslash \{i\},k\in\mathcal{K}}$. Let $\mathbf{b}_{i}^{(n)}\triangleq(\mathbf{e}_{i}^{(n)},\hat{\mathbf{h}}_{-i}^{(n)},\mathbf{e}_{-i}^{(n)})\in\mathcal{B}_{i}$ be the $n$-th batch sample of AP $i$. Also, we denote $\mathbf{p}_{-i}\triangleq\{p_{k,j}\}_{j\in\mathcal{M}\backslash \{i\},k\in\mathcal{K}}$ as the transmit power of other APs. Then, each AP $i$ can estimate the SAA of the ergodic sum-rate \eqref{eq:problem-ergodic-object}, denoted by $\bar{R}_{i}(\mathbf{p}_{i},\mathbf{p}_{-i})$, as
\begin{align}\label{eq:Rbar}
\bar{R}_{i}(\mathbf{p}_{i},\mathbf{p}_{-i})
=\frac{1}{|\mathcal{B}_{i}|}\sum\nolimits_{k\in\mathcal{K}}\sum\nolimits_{\mathbf{b}^{(n)}_{i}\in\mathcal{B}_{i}}\log_{2}(1+\bar{\gamma}_{k,i}(\hat{\mathbf{h}}_{i},\mathbf{b}_{i}^{(n)},\mathbf{p}_{i},\mathbf{p}_{-i})),
\end{align}
where $\bar{\gamma}_{k,i}(\hat{\mathbf{h}}_{i},\mathbf{b}_{i}^{(n)},\mathbf{p}_{i},\mathbf{p}_{-i})$, which is defined in \eqref{eq:gamma_bar} on the top of the next page, stands for the SINR evaluated at AP $i$ over its a batch sample $\mathbf{b}_{i}^{(n)}$. In \eqref{eq:Rbar}, the unknown statistics such as the CSI errors $\mathbf{e}_{i}$ and other-AP information $\hat{\mathbf{h}}_{-i}$ and $\mathbf{e}_{-i}$ are averaged over the mini-batch samples. 
By doing so, we can approximate the ergodic sum-rate with the mini-batch-based SAA. After evaluating $\bar{R}_{i}(\mathbf{p}_{i},\mathbf{p}_{-i})$, each AP $i$ individually performs the projected SGD update as
\begin{align}\label{eq:CSGD}
    \mathbf{p}_{i}\leftarrow \mathcal{P}\left[\mathbf{p}_{i}+\alpha\nabla_{\mathbf{p}_{i}}\bar{R}_{i}(\mathbf{p}_{i},\mathbf{p}_{-i})\right],
\end{align}
where $\nabla_{X}$ denotes the gradient operator with respect to a variable $X$, $\alpha>0$ stands for a step size for each update and $\mathcal{P}[\cdot]$ indicates the projection operator onto the feasible set \eqref{eq:power}. In \eqref{eq:CSGD}, the other-AP solution $\mathbf{p}_{-i}$, which can be informed from other APs, is simply regarded as a constant.

We summarize the CSGD method in Algorithm 1. At each iteration, the APs first exchange their local statistics, i.e., the local long-term CSI $\boldsymbol{\rho}_i$ and local solution $\mathbf{p}_{i}$, through fronthaul channels with the help of the CP. Then, each AP updates its power control vector $\mathbf{p}_{i}$ based on \eqref{eq:CSGD}. This procedure is repeated until the convergence. The CSGD can be realized in a decentralized manner since the update rule \eqref{eq:CSGD} can be executed only with the local information. The computational complexity of the CSGD method is dominated by the gradient calculation. To obtain $\nabla_{\mathbf{p}_i}\bar{R}_i(\mathbf{p}_i,\mathbf{p}_{-i})$, each AP $i$ needs $\mathcal{O}(|\mathcal{B}_{i}|MK^{2})$ calculations since $|\mathcal{B}_{i}|$ batch samples are leveraged for the SAA in \eqref{eq:Rbar}. Defining $L$ as the number of the CSGD iterations, the overall complexity becomes $\mathcal{O}(L|\mathcal{B}_i|MK^2)$.

The iterative nature of the CSGD algorithm not only incurs the prohibitive fronthaul uses for sharing $\mathbf{p}_{i}$ repeatedly but also increases the computational complexity.
To address this challenge, we propose a DL-based approach which only needs one-shot coordination.

\begin{figure*}
\begin{align}\label{eq:gamma_bar}
&\bar{\gamma}_{k,i}(\hat{\mathbf{h}}_{i},\mathbf{b}_{i}^{(n)},\mathbf{p}_{i},\mathbf{p}_{-i}) \!= \\ \notag &\!\frac{\left|
            (\hat{h}_{k,i}\!+\!e_{k,i}^{(n)}) \sqrt{p_{k,i}}\hat{h}_{k,i}^*/|\hat{h}^*_{k,i}| \!+\! \sum_{j\in\mathcal{M}\setminus{\{i\}}}(\hat{h}_{k,j}^{(n)} \!+\! e_{k,j}^{(n)}) \sqrt{p_{k,j}}\hat{h}_{k,j}^{*^{(n)}}/|\hat{h}_{k,j}^{*^{(n)}}|\right|^2}{  1\!+\! \sum_{l\in\mathcal{K}\setminus{\{k\}}}\left|
            (\hat{h}_{k,i}\!+\!e_{k,i}^{(n)}) \sqrt{p_{l,i}}\hat{h}_{l,i}^{*^{(n)}}/|\hat{h}_{l,i}^{*^{(n)}}| \!+\!  \sum_{j\in\mathcal{M}\setminus{\{i\}}}(\hat{h}_{k,j}^{(n)} \!+\! e_{k,j}^{(n)}) \sqrt{p_{l,j}}\hat{h}_{l,j}^{*^{(n)}}/|\hat{h}_{l,j}^{*^{(n)}}|\right|^2}
\end{align}
\end{figure*}


\setlength{\textfloatsep}{7pt}{
\begin{algorithm}[]
\begin{algorithmic}[1]
\caption{CSGD algorithm}\label{alg:cooperative-SGD}
\REPEAT
\STATE Each AP $i$ shares its local solution $\mathbf{p}_{i}$ to the other APs $\mathcal{M}\setminus\{i\}$ via fronthaul links.
\STATE Each AP $i$ samples a mini-batch set from \eqref{eq:sample} and evaluates $\bar{R}_{i}(\mathbf{p}_{i},\mathbf{p}_{-i})$ in \eqref{eq:Rbar}.
\STATE Each AP $i$ updates its local solution $\mathbf{p}_{i}$ from \eqref{eq:CSGD}.
\UNTIL{convergence}
\end{algorithmic}
\end{algorithm}
}

\section{Proposed Cooperative Learning Method\label{sec:Proposed-DL}}

This section proposes a CL framework which optimizes transmit power control variables by means of decentralized CP-AP cooperation. Both the CP and APs are equipped with their own DNN modules which execute the coordination and computation of dedicated entities. We first design a cooperative DNN inference, and it is followed by the joint training strategy.

\begin{figure*}
\centering\includegraphics[width=1\linewidth]{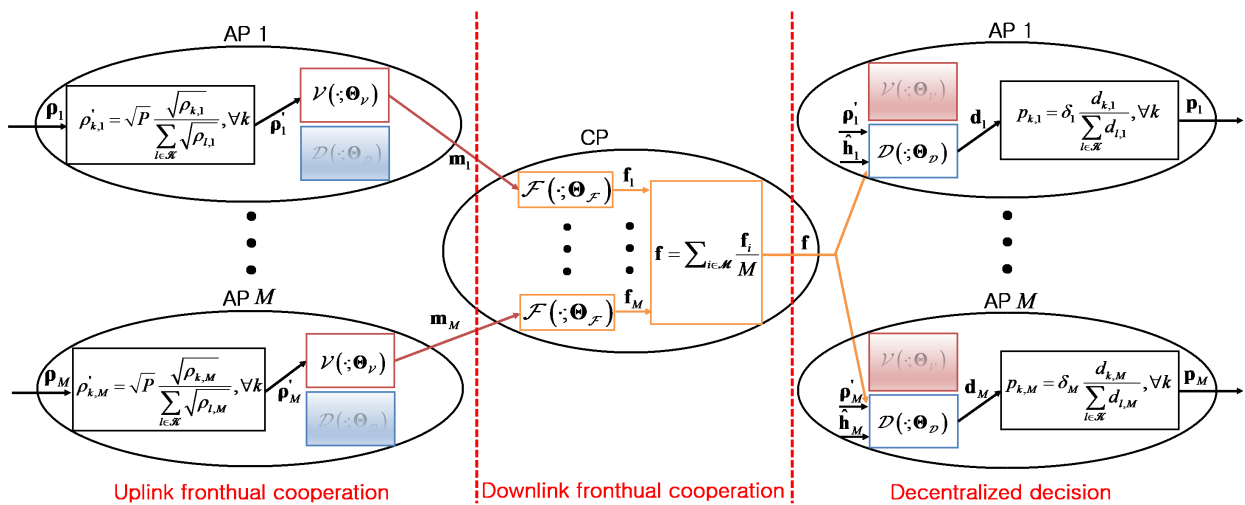}
\caption{{\label{fig:CooperativeStructure} Proposed cooperative learning structure}}
\end{figure*}

\subsection{Cooperative Learning Mechanism}
As shown in Fig. \ref{fig:CooperativeStructure}, the proposed CL framework consists of three sequential steps: uplink fronthaul cooperation, downlink fronthaul cooperation, and decentralized decision. The details of each step are presented in the following.

\subsubsection{Uplink Fronthaul Cooperation}

For the collaborative optimization, the CP needs to collect local information from the APs. To this end, each AP $i$ first conveys an uplink message $\mathbf{m}_{i} \in \mathbb{R}^{d_U}$ of length $d_{U}$ to the CP through the fronthaul link. It is created based on the local long-term CSI $\boldsymbol{\rho}_{i}$ as
\begin{align}
    \mathbf{m}_{i} = \mathcal{V}_i( \boldsymbol{\rho}^{\prime}_{i} ; \boldsymbol{\Theta}_{\mathcal{V}_i} ), \label{eq:uplink-fronthaul-message-DNN}
\end{align}
where $\mathcal{V}_{i}(\cdot;\boldsymbol{\Theta}_{\mathcal{V}_{i}})$ indicates a message-generating DNN at AP $i$ with trainable parameter $\boldsymbol{\Theta}_{\mathcal{V}_{i}}$. The input feature $\boldsymbol{\rho}_i^{\prime}\triangleq\{\rho_{k,i}^{\prime}\}_{k\in\mathcal{K}}$ of the DNN $\mathcal{V}_{i}(\cdot;\boldsymbol{\Theta}_{\mathcal{V}_{i}})$ is obtained as
\begin{align}
    \rho_{k,i}^{\prime} = \sqrt{P} \frac{\sqrt{\rho_{k,i}}}{\sum\nolimits_{l\in\mathcal{K}} \sqrt{\rho_{l,i}}}. \label{eq:pre-processing}
\end{align}
The data pre-processing in \eqref{eq:pre-processing} normalizes the long-term local CSI $\boldsymbol{\rho}_{i}$ so that the resulting input feature $\boldsymbol{\rho}_i'$ lies in a bounded region $\rho^{\prime}_{k,i}\in[0,\sqrt{P}]$. This restricts the DNN inputs into a compact set, thereby satisfying the necessary condition for the universal approximation theorem \cite{Hornik-et-al:NN89}. As a result, the training process can be accelerated by making the DNN focus on the importance of individual channel links $\rho_{k,i}$, but not on their absolute values. The normalized long-term CSI helps the optimization of the cell-free massive MIMO systems \cite{Interdonato-et-al:ICC19,Zaher-et-al:arXiv21}. 

As in \eqref{eq:uplink-fronthaul-message-DNN}, we may leverage individual DNNs $\mathcal{V}_{i}(\cdot;\boldsymbol{\Theta}_{\mathcal{V}_{i}})$ dedicated to each AP $i\in\mathcal{M}$. However, this approach lacks the flexibility to the number of the APs $M$. A group of the DNNs $\{\mathcal{V}_{i}(\cdot;\boldsymbol{\Theta}_{\mathcal{V}_{i}})\}_{i\in\mathcal{M}}$ trained at a certain $M$ cannot be directly applied to other network configurations with different $M$. This requires additional training steps to prepare multiple DNN instances for all possible cell-free massive MIMO setups. Such a scenario prevails in dynamic cell-free massive MIMO systems where only a subset of APs becomes active to achieve the energy efficient communications \cite{Na-et-al:CCNC18}. To resolve this issue, we adopt a scalable learning architecture where the calculations of the DNNs are independent of the number of the APs $M$. We reuse the identical DNN $\mathcal{V}(\cdot;\boldsymbol{\Theta}_{\mathcal{V}})$ to realize the uplink message-generating inference in \eqref{eq:uplink-fronthaul-message-DNN} for all APs as
\begin{align}
    \mathbf{m}_{i} = \mathcal{V}( \boldsymbol{\rho}^{\prime}_{i} ; \boldsymbol{\Theta}_{\mathcal{V}} ). \label{eq:mi}
\end{align}

Such a parameter sharing approach has been widely adopted in existing DNN architectures, e.g., convolutional neural networks (CNNs) and graph neural networks. This is particularly beneficial to improve the generalization ability of the DNNs to work well with unseen input distributions, e.g., with a new system size $M$ unavailable during the training.

\subsubsection{Downlink Fronthaul Cooperation}
The CP aggregates partitioned information vectors $\{\mathbf{m}_{i}\}_{i\in\mathcal{M}}\in\mathbb{R}^{Md_{U}}$ using its DNN $\mathcal{F}(\cdot ; \boldsymbol{\Theta}_{\mathcal{F}})$ with a parameter set $\boldsymbol{\Theta}_{\mathcal{F}}$. The corresponding output, denoted by $\mathbf{f}\in\mathbb{R}^{d_D}$ of length $d_{D}$, acts as a downlink communication message to be broadcasted to all APs. One naive approach for this task is to exploit the concatenation of all uplink messages as an input feature, i.e., $\mathbf{f}=\mathcal{F}(\{\mathbf{m}_{i}\}_{i\in\mathcal{M}} ; \boldsymbol{\Theta}_{\mathcal{F}})$ \cite{Lee-et-al:TWC21}. However, this approach fails to achieve the scalability to the AP population $M$ since the input dimension increases with $M$.

To address this issue, we build a dimensionality-invariant inference at the CP where the computation processes of the downlink message $\mathbf{f}$ become independent with $M$. We first extract a latent feature $\mathbf{f}_{i}\in\mathbb{R}^{d_{D}}$ of the uplink message $\mathbf{m}_{i}$~as
\begin{align}
    \mathbf{f}_{i} = \mathcal{F}\left( \mathbf{m}_{i} ; \boldsymbol{\Theta}_{\mathcal{F}} \right). \label{eq:downlink-fronthaul-message-DNN}
\end{align}
Likewise \eqref{eq:uplink-fronthaul-message-DNN}, the parameter sharing concept is also utilized in \eqref{eq:downlink-fronthaul-message-DNN} so that a sole DNN $\mathcal{F}(\cdot ; \boldsymbol{\Theta}_{\mathcal{F}})$ produces a group of information vectors $\{\mathbf{f}_{i}\}_{i\in\mathcal{M}}$ in parallel. By leveraging the concept of the superposition coding of the non-orthogonal multiple access system, the downlink message vector $\mathbf{f}$ is designed as the average of $\mathbf{f}_{i}$, $\forall i\in\mathcal{M}$, as
\begin{align}\label{eq:Downlnk-message-aggregation}
\mathbf{f}=\frac{1}{M}\sum\nolimits_{i\in\mathcal{M}}\mathbf{f}_{i}=\frac{1}{M}\sum\nolimits_{i\in\mathcal{M}}\mathcal{F}\left( \mathbf{m}_{i} ; \boldsymbol{\Theta}_{\mathcal{F}} \right).
\end{align}
Notice that \eqref{eq:Downlnk-message-aggregation} is viewed as the average pooling operation which has been widely adopted for CNNs. This operation is beneficial to extract an important global feature $\mathbf{f}$ from local AP message vectors $\{\mathbf{f}_{i}\}_{i\in\mathcal{M}}$ by pruning unnecessary statistics. As a consequence, we can facilitate the dimensionality-invariant fronthaul cooperation efficiently.

\subsubsection{Decentralized Decision}

To determine the local power allocation solution $\mathbf{p}_{i}$, each AP $i$ utilizes the downlink message $\mathbf{f}$ received from the CP along with its local CSI $\boldsymbol{\rho}'_i$ and $\hat{\mathbf{h}}_i$. Such a decentralized decision-making process at all APs is modeled by a DNN $\mathcal{D}(\cdot ; \boldsymbol{\Theta}_{\mathcal{D}})$ with trainable parameter $\boldsymbol{\Theta}_{\mathcal{D}}$. As a result, each AP $i$ recovers its solution $\mathbf{p}_{i}$ as
\vspace{-1mm}
\begin{align}
    \mathbf{p}_{i} = \mathcal{D}(\mathbf{f}, \boldsymbol{\rho}'_{i}, \hat{\mathbf{h}}_{i};\boldsymbol{\Theta}_{\mathcal{D}}). \label{eq:downlink-decision-DNN}
\end{align}
Since the output of $\mathcal{D}(\cdot;\boldsymbol{\Theta}_{\mathcal{D}})$ is directly exploited as the transmit power variables, an appropriate design for the output layer is necessary to guarantee the power constraint \eqref{eq:power}. To this end, we develop a novel activation function at the output layer. Let $\mathbf{d}_i = [d_{1,i} \cdots d_{K,i}, \delta_{i}] \in\mathbb{R}^{K+1}$ be the output vector of $\mathcal{D}_{i}(\cdot, \boldsymbol{\Theta}_{\mathcal{D}})$ before the activation function. The first $K$ elements $d_{k,i}\geq 0$, $k\in\mathcal{K}$ control ratios among the transmit power variables $p_{k,i}$, $k\in\mathcal{K}$. On the contrary, the last element $\delta_{i}$ determines the total transmit power to be consumed by AP $i$. To restrict $\delta_{i}$ into the feasible range $[0,P]$, we apply the rectified linear unit 6 (ReLU6) function as
\vspace{-1mm}
\begin{align}
    \delta_{i}\leftarrow P \min\left( \max\left( \delta_{i}, 0 \right), 6 \right)/ 6. \label{eq:scaled-ReLU6}
\end{align}
The power control variable $p_{k,i}$ is then retrieved as
\begin{align}
    p_{k,i} = \frac{\delta_i d_{k,i}} {\sum\nolimits_{l\in\mathcal{K}} d_{l,i}}, \label{eq:scaling-power-control}
\end{align}
which always leads to the feasible solution as $\sum_{k\in\mathcal{K}} p_{k,i} = \delta_i\leq P$. The output activation of the DNN $\mathcal{D}(\cdot ; \boldsymbol{\Theta}_{\mathcal{D}})$ can be specified by the operations in \eqref{eq:scaled-ReLU6} and \eqref{eq:scaling-power-control}. 

Finally, a group of DNNs in \eqref{eq:mi}, \eqref{eq:Downlnk-message-aggregation}, and \eqref{eq:downlink-decision-DNN} provides an end-to-end forward pass mapping $\mathcal{G}(\cdot;\boldsymbol{\Theta})$ of the proposed CL as $\mathbf{p} = \mathcal{G}( \boldsymbol{\rho}, \hat{\mathbf{h}}; \boldsymbol{\Theta} )$, 
where $\boldsymbol{\Theta} \triangleq \{\boldsymbol{\Theta}_{\mathcal{V}}, \boldsymbol{\Theta}_{\mathcal{F}}, \boldsymbol{\Theta}_{\mathcal{D}}\} $ indicates the collection of all trainable parameters.

\vspace{-4mm}

\subsection{Training and Implementation}

We discuss a joint training process of the proposed CL architecture. Plugging $\mathbf{p} = \mathcal{G}( \boldsymbol{\rho}, \hat{\mathbf{h}}; \boldsymbol{\Theta} )$ into (\ref{eq:problem-ergodic}) leads to the training problem written by
\begin{align}
&\underset{\boldsymbol{\Theta}}{\mathrm{max.}}\,\, \mathbb{E}_{ \hat{\mathbf{h}}, \mathbf{e}, \boldsymbol{\rho} } \left[ \sum_{k\in\mathcal{K}} \log_2\left( 1 + \gamma_k\left(\hat{\mathbf{h}}, \mathbf{e}, \mathcal{G}( \boldsymbol{\rho}, \hat{\mathbf{h}}; \boldsymbol{\Theta} )\right)\right) \right], \label{eq:problem-DNN-ergodic}
\end{align}
where the power constraint \eqref{eq:power} is removed since it is always satisfied by \eqref{eq:scaled-ReLU6} and \eqref{eq:scaling-power-control}. The training problem in \eqref{eq:problem-DNN-ergodic} can be readily addressed by existing mini-batch SGD algorithms, e.g., the Adam optimizer. A training dataset contains numerous realizations of the long-term CSI $\boldsymbol{\rho}$. At each training epoch, we randomly sample a mini-batch set consisting of the long-term CSI $\boldsymbol{\rho}$. These can be collected in advance from the experiments or can also be generated based on the known AP-UE deployment scenarios. We then generate the CSI estimate $\hat{\mathbf{h}}$ and error vector $\mathbf{e}$ using the known distributions \eqref{eq:sample}. Since the error ratio $\phi$ randomly changes in practice, it is necessary to build versatile DNNs adaptive to arbitrarily varying $\phi$. To this end, we randomly create the error ratio factor in the training step, e.g., from the uniform distribution $\phi\sim\mathcal{U}(0,1)$. As a result, the proposed CL can be universally applied to any CSI error statistics $\phi$. These are leveraged to calculate the gradient of the training objective \eqref{eq:problem-DNN-ergodic} averaged over the mini-batch set. As a result, the proposed CL is trained by observing a number of artificially generated CSI error samples. By doing so, the DNNs can learn the unknown distribution of the actual CSI based on its estimates, thereby producing a robust power control mechanism.

The proposed joint training process is implemented in an offline manner by collecting all component DNNs at the CP. Trained DNN modules are installed at desired network entities for the decentralized power optimization. In this implementation stage, we no longer need the CSI error $\mathbf{e}$ since the proposed CL $\mathcal{G}( \boldsymbol{\rho}, \hat{\mathbf{h}}; \boldsymbol{\Theta} )$ only accepts the long-term CSI $\boldsymbol{\rho}$ and the estimate of short-term CSI $\hat{\mathbf{h}}$.

The number of the APs $M$ can be regarded as a hyperparameter of the proposed CL strategy. Let $M_{\text{train}}$ be the AP population employed in the training step. To further improve the scalability, $M_{\text{train}}$ should be chosen carefully such that the proposed CL trained at a specific $M_{\text{train}}$ can work well universally over a wide range of the test AP population $M_{\text{test}}$. 
This issue will be clarified in Sec. \ref{sec:Numerical-Results}.

\section{Numerical Results\label{sec:Numerical-Results}}

This section validates the proposed CL method. The APs and UEs are uniformly distributed within a circular area of radius 300 m. The long-term CSI is modeled as $\rho_{k,i}=P_0(q_{k,i}/q_0)^{-\eta}$, where $q_{k,i}$ is the distance between AP $i$ and UE $k$, $P_0=10$ is the path-loss at the reference distance $q_0=30$ m, and $\eta=3$ denotes the path-loss exponent. All component DNNs are realized with $16$ fully-connected hidden layers. The output dimensions of all hidden layers are set to $160K$. The ReLU activation is employed at all hidden layers, which is followed by the batch-normalization layers. The message dimensions are fixed to $d_U = d_D = K$. The performance of the trained DNNs are evaluated over $2\times10^5$ test samples.

We consider the following benchmark schemes. 
\begin{itemize}
    \item \textit{CSGD algorithm:} The CSGD method in Algorithm \ref{alg:cooperative-SGD} generates an unachievable upper bound performance for the proposed CL framework since it allows the APs to share the power control solutions $\mathbf{p}_{i}$, $\forall i\in\mathcal{M}$, iteratively.
    \item \textit{Non-cooperative learning (NCL)}: This scheme does not allow any message exchange among APs, i.e., $d_U = d_D = 0$, thereby presenting a lower bound performance of the proposed coordination mechanism.
    \item \textit{Synthetic CL (SCL) \cite{Zaher-et-al:arXiv21}:} This approach relies on a man-made cooperation strategy. The uplink message $\mathbf{m}_{i}$ is simply fixed as the local long-term CSI as $\mathbf{m}_{i}=\boldsymbol{\rho}_i$. Also, the $k$-th element $f_{k,i}$ of the downlink message vector $\mathbf{f}_{i}$ is given by the normalized local long-term CSI, i.e., $f_{k,i}=\sqrt{P\rho_{k,i}/\Sigma_{j\in\mathcal{M}}\rho_{k,j}}$.
    \item {\textit{Equal power allocation:}} Each AP $i$ employs equally allocated power $p_{k,i}=P/K$, $\forall k\in\mathcal{K}$, for all UEs.
\end{itemize}
Notice that the SCL baseline in \cite{Zaher-et-al:arXiv21} does not apply the parameter sharing policy, i.e., individual APs have dedicated DNN modules. Thus, these baselines should prepare multiple instances of DNNs trained at each of AP populations $M$. On the contrary, the proposed scalable method exploits a single set of DNNs at all possible $M$.

\begin{figure}

\begin{subfigure}[b]{1\textwidth}
\includegraphics[width=0.8\linewidth]{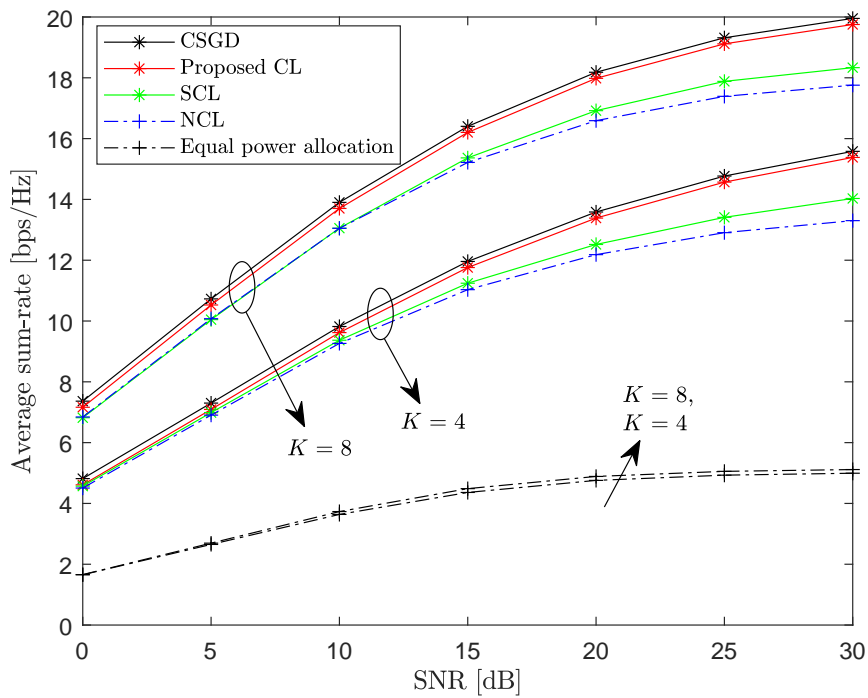}
\centering\caption{{\label{fig_vsSNR1}Single-antenna AP, $M=8$}}
\end{subfigure}

\begin{subfigure}[b]{1\textwidth}
\includegraphics[width=0.8\linewidth]{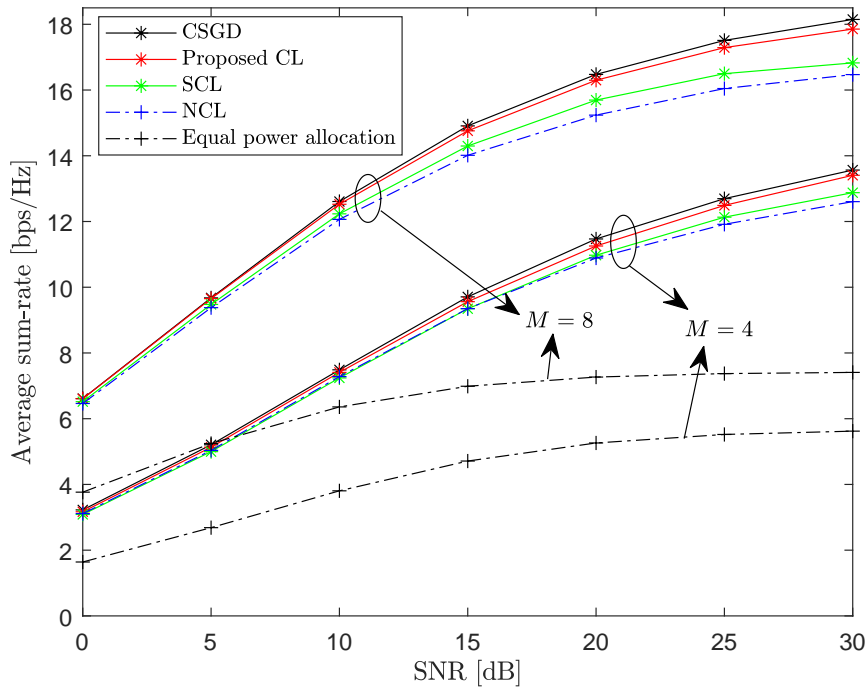}
\centering\caption{{\label{fig_vsSNR2}Multi-antenna AP, $N=2$, $K=4$}}
\end{subfigure}

\caption{{\label{fig_vsSNR}Average sum-rate versus SNR with $M\in\{4,8\}$, $K\in\{4,8\}$ and $\phi=0.1$.}}

\end{figure}

Fig. \ref{fig_vsSNR} depicts the average sum-rate performance by varying the SNR which is defined as $P$. Fig. 2(a) shows that the proposed CL method exhibits negligible loss compared to the CSGD algorithm. This validates the effectiveness of the proposed approach for achieving an upper bound performance. The proposed CL is superior to the DL baselines for all simulated scenarios. The performance gain becomes significant in the high SNR regime where the inter-user interference dominates the overall system capacity. This infers that the proposed fronthaul coordination strategy is crucial for the CP to mitigate the interference.

Fig. 2(b) plots the average sum-rate performance for the multi-antenna AP scenario, in
which each AP has two antennas. The proposed CSGD and CL methods can be straightforwardly applied to this multi-antenna AP scenario with a slight modification on the transmit signal $x_i$ in (3). Here, we have vector-valued signal $\mathbf{x}_{i}\in\mathbb{C}^{N}$ with $N$ AP antennas, where $\mathbf{x}_i = \sum_{k\in\mathcal{K}}\sqrt{p_{k,i}}\hat{\mathbf{h}}_{k,i} s_k / ||\hat{\mathbf{h}}_{k,i}||$ with $\hat{\mathbf{h}}_{k,i}\in\mathbb{C}^{N}$ being the estimated CSI vector from AP $i$ to UE $k$. From the figure, it is seen that the proposed CL still performs well with the multi-antenna APs, showing an almost identical performance to the CSGD upper bound. Thus, we can conclude that the proposed CL can be applied to an arbitrary number of AP antennas.

\begin{figure}
\centering\includegraphics[width=0.80\linewidth]{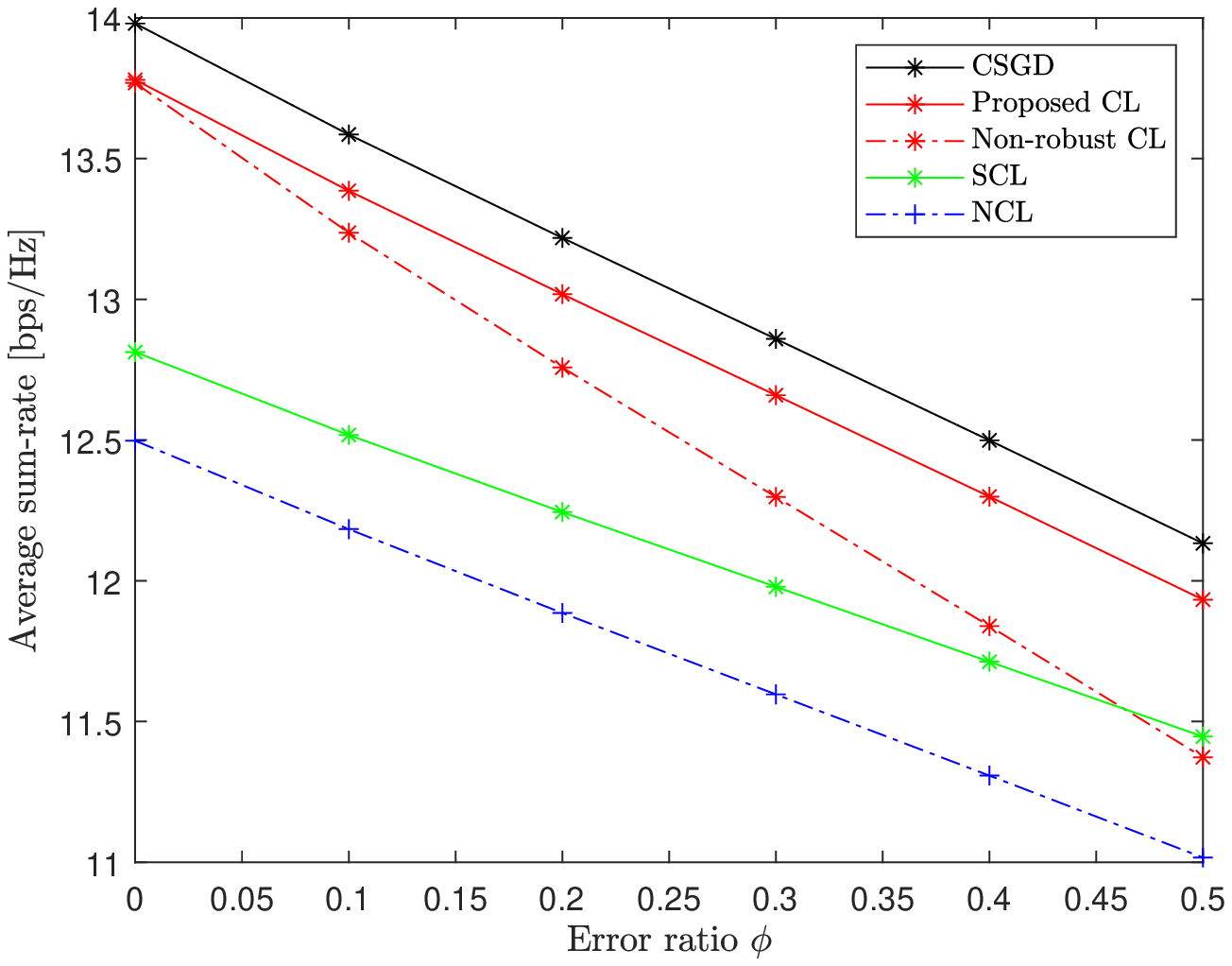}
\caption{{\label{fig_vsER1}Average sum-rate versus error ratio $\phi$ for $M=8$, $K=4$ and SNR = 20 dB}}
\end{figure}

Fig. \ref{fig_vsER1} plots the average sum-rate versus error ratio $\phi$ for $M=8$, $K=4$ and SNR = 20 dB. To validate the robustness of the proposed approach, in Fig. \ref{fig_vsER1}, we also assess a non-robust method where the DNNs are trained only with the actual channels, i.e., the CSI estimation error $\mathbf{e}$ is not included in the training process. The proposed scheme shows superior performance than the other schemes, while exhibiting negligible loss to the CSGD algorithm. The gain achieved by the proposed CL method becomes pronounced as the error ratio gets larger.
The proposed CL includes the fronthaul coordination into the optimization domain. Thus, the optimized fronthaul messages act as sufficient statistics for the decentralized decision DNN at the APs. However, for a large CSI error $\phi=0.5$, the SCL performs better than the non-robust CL, implying the importance of the robust training policy in the high CSI error regime.

\begin{figure}
\centering\includegraphics[width=0.80\linewidth]{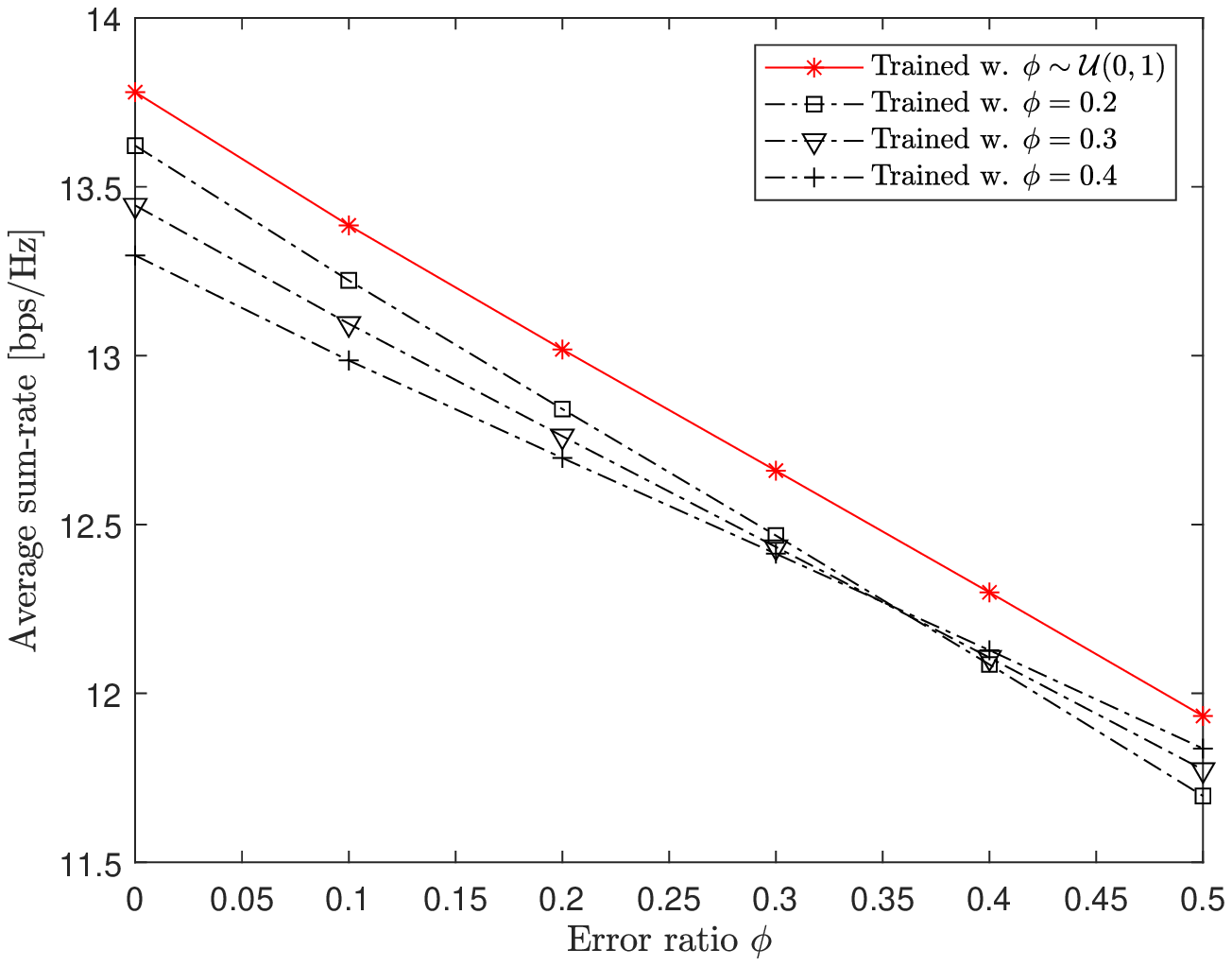}
\caption{{\label{fig_vsER2}Average sum-rate versus error ratio $\phi$ for $M=8$, $K=4$ and SNR = 20 dB}}
\end{figure}

Fig. \ref{fig_vsER2} depicts the average sum-rate performances of the proposed CL according to the sampling policy of the error ratio $\phi$ for $M=8$, $K=4$ and SNR = 20 dB. From the figure, the sum-rate performance is degraded when the training $\phi$ differs from the test $\phi$. From the result, we can know that training with accurate $\phi$ is important for the robust optimization of the proposed CL. Also, the proposed CL trained with randomly sampled $\phi$ from $\phi\sim\mathcal{U}(0,1)$ shows improved performance than that trained with fixed $\phi$ even at the perfectly known error ratio region. This result implies that our robust training policy, which randomly samples $\phi$, not only prevents the performance degradation from the absence of perfect long-term statistics in the online phase but also can improve the performance by securing a regularization effect such as the noise injection \cite{Goodfellow-et-al:IMT16}.

\begin{table}[] \centering
\caption{{\label{table:scalability2}}Relative sum-rate for $K=16$, $\phi=0.1$ and $\text{SNR}=20\ \text{dB}$}
\renewcommand{\tabcolsep}{1.2mm}
\begin{tabular}{|c|c|c|c|c|c|}
\hline
\multicolumn{1}{|c|}{\diagbox[]{$M_{\text{train}}$}{$M_{\text{test}}$}} & 16 & 20 & 24 & 28 & 32 \\ \hline
8 & 0.93 & 0.93 & 0.92 & 0.94 & 0.92 \\ \hline
16 & 0.94 & 0.95 & 0.94 & 0.96 & 0.96 \\ \hline
24 & 0.93 & 0.96 & 0.99 & 0.97 & 0.97 \\ \hline
32 & 0.93 & 0.97 & 0.97 & 0.98 & 0.99 \\ \hline
\end{tabular}
\end{table}

Next, we investigate the flexibility and scalability of the proposed CL method by employing different numbers of the APs $M$ in the training and testing steps. To be specific, the proposed CL trained at $M_{\text{train}}$ APs is straightforwardly applied to the cell-free massive MIMO system with $M_{\text{test}}$ APs. We evaluate the relative sum-rate performance, which is defined as the sum-rate achieved by the proposed scheme normalized by that of the CSGD method. Table \ref{table:scalability2} lists the relative sum-rate performance of the proposed CL trained with $M_{\text{train}}\in\{8,16,24,32\}$ and tested with $M_{\text{test}}\in\{16,20,24,28,32\}$ for $K=16$, $\phi=0.1$ and SNR = 20 dB. The proposed CL generally performs well overall simulated $M_{\text{test}}$ even in the massive AP regime with $M=32$ although other configurations $M_{\text{test}}\neq M_{\text{train}}$ were not observed in the training step. Therefore, we only need a single training process for the optimized hyperparameter $M_{\text{train}}$. This results in a huge reduction in the training complexity compared to existing methods \cite{Zaher-et-al:arXiv21,Lee-et-al:TWC21} which should be trained at all possible $M_{\text{test}}$.

\section{Conclusions\label{sec:Conclusion}}
This paper has presented a DL-based decentralized power control strategy for cell-free massive MIMO systems. A novel CL structure has been proposed which realizes fronthaul cooperation and decentralized decision processes using component DNNs. As a consequence, the CP and APs can identify an appropriate power control mechanism by exchanging DNN-oriented communication messages.
The notion of the parameter sharing has been employed so that the forwardpass computations of the proposed CL become independent of the AP population. Numerical results have demonstrated the effectiveness of the proposed approach. The extension to the multi-antenna UE scenario is worth to be considered as a future work.


\begin{thebibliography}{15}

\bibitem{Choi-et-al:TWC20}
J. Choi, N. Lee, S.-N. Hong and G. Caire, ``Joint user selection, power allocation and precoding design wtih imperfect CSIT for multi-cell MU-MIMO downlink systems,'' \textit{IEEE Trans. Wireless Commun.}, vol. 19, no. 1, pp. 162--176, Jan. 2020.

\bibitem{Nayebi-et-al:TWC17}
E. Nayebi, A. Ashikhmin, T. L. Marzetta, H. Yang and B. D. Rao, ``Precoding and power optimization in cell-free massive MIMO systems,'' \textit{IEEE Trans. Wireless Commun.}, vol. 16, no. 7, pp. 4445--4459, Jul. 2017.

\bibitem{Ngo-et-al:TWC17}
H. Q. Ngo, A. Ashikhmin, H. Yang, E. G. Larsson and T. L. Marzetta, ``Cell-free massive MIMO versus small cells,'' \textit{IEEE Trans. Wireless Commun.}, vol. 16, no. 3, pp. 1834--1850, Mar. 2017.


\bibitem{Yu-et-al:WCL21}
D. Yu, H. Lee, S.-H. Park and S.-E. Hong, ``Deep learning methods for joint optimization of beamforming and fronthaul quantization in cloud radio access networks,'' \textit{IEEE Wireless Commun. Lett.}, vol. 10, no. 10, pp. 2180--2184, Oct. 2021.

\bibitem{Zaher-et-al:arXiv21}
 M. Zaher, O. T. Demir, E. Bj{\"o}rnson and M. Petrova, ``Learning-based downlink power allocation in cell-free massive MIMO systems,'' \textit{IEEE Trans. Wireless Commun.}, vol. 22, no. 1, pp. 174-188, Jan. 2023.


\bibitem{Lee-et-al:TWC21}
H. Lee, J. Kim and S.-H. Park, ``Learning optimal fronthauling and decentralized edge computation in fog radio access networks,'' \textit{IEEE Trans. Wireless Commun.}, vol. 20, no. 9, pp. 5599--5612, Sep. 2021.

\bibitem{Shamir-Srebro:Allerton14}
O. Shamir and N. Srebro, ``Distributed stochastic optimization and learning,'' in \textit{Proc. Annu. Allerton Conf. Commun. Control Comput. (Allerton)}, Feb. 2015.


\bibitem{Yin-et-al:JSAC13}
H. Yin, D. Gesbert, M. Filippou and Y. Liu, ``A coordinated approach to channel estimation in large-scale multiple-antenna systems,'' \textit{IEEE J. Sel. Areas Commun.}, vol. 31, no. 2, pp. 264--273, Feb. 2013.

\bibitem{Nosrat-Makouei-at-al:TSP11}
B. Nosrat-Makouei, J. G. Andrews and R. W. Heath, ``MIMO interference alignment over correlated channels with imperfect CSI,'' \textit{IEEE Trans. Signal Process.}, vol. 59, no. 6, pp. 2783--2794, Jun. 2011.


\bibitem{Dartmann-et-al:VTC09}
G. Dartmann, M. Jordan, X. Gong and G. Ascheid, ``Intercell interference mitigation with long-term beamforming and low SINR feedback rate in a multiuser multicell unicast scenario,'' in \textit{Proc. IEEE Veh. Technol. Conf. (VTC)}, May 2009.


\bibitem{Riera-Palou-et-al:Allerton19}
F. Riera-Palou and G. Femenias, ``Decentralization issues in cell-free massive MIMO with zero-forcing,'' in \textit{Proc. Annu. Allerton Conf. Commun. Control Comput. (Allerton)}, Sep. 2019.



\bibitem{Ngo-et-al:SPAWC15}
H. Q. Ngo, A. Ashikhmin, H. Yang, E. G. Larsson and T. L. Marzetta, "Cell-free massive MIMO: Uniformly great service for everyone," in \textit{Proc. IEEE Int. Workshop Signal Process. Adv. Wireless Commun. (SPAWC)}, pp. 201--205, Jun. 2015.


\bibitem{Hornik-et-al:NN89}
K. Hornik, M. Stinchcombe and H. White, ``Multilayer feedforward networks are universal approximators,'' \textit{Neural Netw.}, vol. 2, no. 5, pp. 359--366, Jan. 1989.



\bibitem{Interdonato-et-al:ICC19}
G. Interdonato, P. Frenger and E. G. Larsson, ``Scalability aspects of cell-free massive MIMO,'' in \textit{Proc. IEEE Int. Conf. Commun. (ICC)}, May 2019.


\bibitem{Na-et-al:CCNC18}
J. Na, J. Koh, S. Park and J. Kang, ``Energy efficiency enhancement on cloud and edge processing by dynamic RRH selection,'' in \textit{Proc. IEEE Annu. Consum. Commun. Netw. Conf. (CCNC)}, Jan. 2018.




\bibitem{Goodfellow-et-al:IMT16}
I. Goodfellow, Y. Bengio and A. Courville, \textit{Deep Learning}. Cambridge, MA, USA: MIT Press, 2016.






\end{thebibliography}
\end{document}